\documentclass[conference,10pt]{IEEEtran}
\usepackage{graphicx}
\usepackage{amsmath}
\usepackage{mathtools}
\usepackage{amssymb}
\usepackage{math_symbols}
\usepackage{calc}
\usepackage{booktabs,hhline} 
\usepackage{enumitem}
\usepackage{siunitx}
\usepackage{multirow}
\usepackage{acronym}
\usepackage{float}
\usepackage[english]{babel}
\usepackage{times}
\usepackage{url}
\usepackage{algorithm}
\usepackage{algpseudocode}
\usepackage{tikz}
\usetikzlibrary{calc,arrows,positioning}
\usepackage{tabularx}
\acrodef{ssm}[SSM]{state-space model}
\acrodef{ss}[SS]{state-space}
\acrodef{kf}[KF]{Kalman filter} 
\acrodef{pf}[PF]{particle filter} 
\acrodef{bpf}[BPF]{bootstrap \ac{pf}} 
\acrodef{kg}[KG]{Kalman gain}
\acrodef{ekf}[EKF]{extended \ac{kf}}
\acrodef{ckf}[CKF]{cubature \ac{kf}}
\acrodef{ukf}[UKF]{unscented Kalman filter}
\acrodef{brr}[BRR]{Bayesian recursive relation} 
\acrodef{pdf}[PDF]{probability density function}
\acrodef{cke}[CKE]{Chapman-Kolmogorov equation}
\acrodef{gmf}[GMF]{Gaussian mixture filter}
\acrodef{gm}[GM]{Gaussian mixture}
\acrodef{gf}[GF]{Gaussian filter}
\acrodef{gaf}[GAF]{Gaussian-assumed filter}
\acrodef{gpb}[GPB]{generalized pseudo-Bayesian}
\acrodef{imm}[IMM]{interacting multiple-model}
\acrodef{ut}[UT]{unscented transform}
\acrodef{mc}[MC]{Monte Carlo}
\acrodef{rmse}[RMSE]{root mean-squared error}
\acrodef{mse}[MSE]{mean-squared error}
\acrodef{anees}[ANEES]{average normalized estimate error squared}
\acrodef{pmf}[PMF]{point-mass filter}
\acrodef{kld}[KLD]{Kullback-Leibler divergence}
\acrodef{sde}[SDE]{Shannon differential entropy}
\acrodef{ode}[ODE]{ordinary differential equations}
\acrodef{vb}[VB]{variational Bayes}
\acrodef{ai}[AI]{artificial intelligence} 
\acrodef{dnn}[DNN]{deep neural network} 
\acrodef{cnn}[CNN]{convolutional neural network} 
\acrodef{nn}[NN]{neural network}  
\acrodef{gnn}[GNN]{graph neural network} 
\acrodef{rnn}[RNN]{recurrent neural network} 
\acrodef{fc}[FC]{fully connected} 
\acrodef{snr}[SNR]{signal-to-noise ratio}
\acrodef{ml}[ML]{machine learning}
\acrodef{gru}[GRU]{gated recurrent unit} 
\acrodef{lstm}[LSTM]{long short-term memory} 
\acrodef{rkn}[RKN]{recurrent Kalman network}
\acrodef{pbm}[PBM]{physics-based model}
\acrodef{apbm}[APBM]{data-augmented physics-based model}
\acrodef{tm}[TM]{true model}
\acrodef{rts}[RTS]{Rauch-Tung-Striebel}
\acrodef{ddm}[DDM]{data-driven model}
\acrodef{vae}[VAE]{variational autoencoder}
\acrodef{pinn}[PINN]{physics-informed neural network}

\acrodef{narma}[NARMA]{nonlinear autoregressive moving-average}
\acrodef{armax}[ARMAX]{autoregressive moving-average model}
\acrodef{ungm}[UNGM]{univariate nonstationary Gaussian model}

\acrodef{nat}[NaT]{navigation and tracking}

\IEEEoverridecommandlockouts

\newcommand\clearrow{\global\let\rowmac\relax}
\clearrow

\def\bS{\bfSigma}

\def\bSv{\bS^{\bfv}}

\def\bbS{\bar{\bfSigma}}
\def\bbSv{\bbS^{\bfv}}

\def\rmse{\mathrm{RMSE}}
\def\anees{\mathrm{ANEES}}

\def\St{\ensuremath{\operatorname{St}}}
\def\nuv{\nu^{\bfv}}

\def\xpos{\mathsf{x}}
\def\ypos{\mathsf{y}}
\def\tpdf{\mathcal{T}}
\def\gpdf{\mathcal{G}}
\DeclareMathOperator{\atantwo}{atan2}

\begin{document}
    
\title{Fixed-structure Gaussian Mixture Filtering\\with Robust Measurement Updates under Outliers}

\author{%
	\IEEEauthorblockN{Ond\v{r}ej Straka}
	\IEEEauthorblockA{%
		European Centre of Excellence NTIS, \\
		University of West Bohemia in Pilsen, Czech Republic\\ E-mail: straka30@fav.zcu.cz}
	\and
	\IEEEauthorblockN{Uwe D. Hanebeck}
	\IEEEauthorblockA{Intelligent Sensor-Actuator-Systems Laboratory (ISAS),\\ Karlsruhe Institute of Technology Karlsruhe, Germany\\ E-mail: uwe.hanebeck@kit.edu}
\thanks{The work was partially supported by the Ministry of Education, Youth and Sports of the Czech Republic under the OP JAC project DigiTech no.
CZ.02.01.01/00/23\_021/0008436.}
}%
\selectlanguage{english}
\maketitle

\begin{abstract}
Bayesian state estimation for discrete-time nonlinear stochastic systems is considered in the presence of measurement outliers.
Building on a fixed-structure Gaussian mixture filtering framework, this paper proposes a robust measurement-update variant in which the predictive density structure is determined by an offline decomposition of the transition density into axis-aligned Gaussian components. 
This construction maintains the Gaussian mixture structure as deterministic and tunable via the chosen decomposition fidelity.
Measurement components affected by outliers are modeled using a Student's-t distribution, and the corresponding update of each Gaussian mixture component is approximated by a variational Bayes procedure.
The resulting filter is evaluated in a two-dimensional tracking scenario with a three-dimensional augmented state and range-and-bearing measurements, where the bearing channel is affected by outliers modeled as heavy-tailed noise.
\end{abstract}
\section{Introduction}
State estimation is crucial for fields such as tracking, guidance, positioning, navigation, sensor fusion, control, fault detection, and decision-making.
Its goal is to estimate the state of a dynamic system from noisy measurements.
The Bayesian approach to state estimation involves calculating the posterior \ac{pdf} of the state conditioned on available measurements. 
For discrete-time problems, it involves alternating between the Bayes equation and \ac{cke}, collectively referred to as \acp{brr}.
\acp{brr} are analytically tractable only for a few special cases, such as linear Gaussian systems, and usually an approximation is required.

The approximate solutions to the \acp{brr} yield algorithms of varying complexity.
Assuming the joint state and measurement prediction \ac{pdf} being Gaussian leads to \acp{gaf} (e.g., cubature filter~\cite{ArHa:09} or the stochastic integration filter~\cite{DuStrSi:13}), which are computationally light.
However, this assumption rarely holds, leading to poor \ac{gf} performance in strongly nonlinear systems.

Discrete approximation of the state posterior \ac{pdf} results in another group of approximate Bayesian algorithms.
These include the \acp{pmf}~\cite{SiKraSo:06} replacing the continuous support of the posterior by a usually orthogonal and equidistant grid of weighted points. While such a grid is flexible in representing the posterior, it leads to computationally intensive algorithms.
Switching from a deterministically constructed grid to randomly positioned points leads to \acp{pf}~\cite{RiArGo:03}.
They are computationally lighter than the \acs{pmf}. Still, the estimates are influenced by random effects, which can pose significant challenges in safety-critical applications or even render them inadmissible in certified solutions.

The gap between \acp{gaf} and \acp{pmf} is filled by \acp{gmf}~\cite{sorensonRecursiveBayesianEstimation1971,StraDuSi:11b}.
They represent the posterior \ac{pdf} as a mixture of Gaussian \acp{pdf} with continuous support.
The support, along with smaller variances of each \ac{gm} term compared to a single Gaussian \ac{pdf}, lead to higher accuracy, even in highly nonlinear problems. 
Compared to \acp{pf} and \acp{pmf}, the \acp{gmf} are generally less computationally intensive.

Many \acp{gmf} have been proposed, varying in the way how individual components in the \ac{gm} are handled. \emph{Local} \acp{gmf}~\cite{SoAl:71} process each Gaussian component with its own filter and control the inevitable growth of components by pruning or merging, an approach later refined by~\cite{ItoXi:00} using a bank of \acp{gaf} with rules for updating component weights after correction.
More recent work~\cite{AAS25_Durant} further refines post-correction weight updates using various linearization strategies, while~\cite{kotechaGaussianSumParticle2003} employ more sophisticated component-wise filtering through a bank of Gaussian particle filters.

\emph{Global} \acp{gmf} process all mixture components jointly, beginning with~\cite{CDC03_Feiermann-ProgBayes} posterior parameters are computed via \acp{ode} instead of approximating the prior mixture, and more recently, including~\cite{FUSION24_Frisch}, where a deterministic-sample update by weighting prior samples is performed, extracting higher-order moments, and deriving a posterior \ac{gm} that minimizes Fisher information.
Another line of global methods uses \ac{gm} approximations of the transition density (e.g., \cite{willsNumericallyRobustBayesian2023}), which generally causes exponential growth of components and hence requires complex mixture reduction, whereas in~\cite{MFI06_Huber} this was avoided by decomposing the transition density into axis-aligned components, enabling a fixed number of components and closed-form posterior expressions.

Recently, \cite{Straka25:Efficient} proposed an efficient \emph{fixed-structure} \acp{gmf} that avoid local component processing and prevent component explosion by decomposing the transition density offline into axis-aligned Gaussian components. Two decompositions were introduced: the decomposition of the \emph{filtered state grid} and of the \emph{predicted state grid}, each offering different trade-offs among offline complexity, the need for an approximation domain, and the use of numerical integration. Both decompositions maintain a fixed number of posterior components, enabling a tunable balance between online computational load and estimation accuracy.

Measurement noise often contains outliers caused by unmodeled anomalies or sensor faults, making Gaussian noise assumptions unreliable and motivating the use of heavy-tailed models, such as the Student's-t distribution. Existing outlier-robust state estimation methods include PDF-based approaches like Gaussian sum, \acp{pmf}, and \acp{pf}; robust \acp{gaf} that overbound noise covariances (often degrading accuracy); and Student's-t–based filters~\cite{Straka2017Stochastic} that offer solid theoretical grounding with low computational cost. Recent advances in the latter group include variational methods~\cite{AgNiNe:12}, Student's-t process regression~\cite{SoSa:15}, and solutions to \acp{brr} for linear and nonlinear systems with heavy-tailed or skewed Student's-t noise \cite{RoOzGu:13,TroHoSa:16}.

This paper builds upon the fixed-structure \ac{gmf} with predicted-state grid decomposition, which offers easier utilization in higher-dimensional state spaces and greater flexibility in the \emph{denseness} of the Gaussian terms and state-space coverage compared to the filtered-state grid.
Here, denseness refers to the spatial concentration of mixture terms in the approximation region, not to a probability density function.
A robust \emph{fixed-structure \ac{gmf}} is proposed to address the Bayesian state estimation problem involving outliers in measurements.
The novelty lies in the combination of the predicted state grid fixed structure with robust heavy-tailed update.

The paper is structured as follows: Section~\ref{sec:problem_Statement} specifies the estimation problem to be addressed. The \acp{brr} and the fixed-structure \ac{gmf} are given in Section~\ref{sec:structured_gmf}. The robust measurement update of the filter addressing the outliers is presented in Section~\ref{sec:measurement_update}. Performance of the proposed filter is demonstrated using a 3D tracking example in Section~\ref{sec:numerical illustration}, and concluding remarks are drawn in Section~\ref{sec:conclusion}.
\section{Problem Statement}\label{sec:problem_Statement}
Consider a discrete-time stochastic system described by a nonlinear state-space model
\begin{subequations}\label{eq:sseq}
\begin{align}\label{eq:sseqx}
    \bfx_{k+1} &= \bff_k(\bfx_k) + \bfw_k,\\
    \bfz_k &= \bfh_k(\bfx_k) + \bfv_k,\label{eq:sseqz}
\end{align}
\end{subequations}
where $\bfx_k\in\real^{n_x}$ and $\bfz_k\in\real^{n_z}$ represent the immeasurable state of the system and the measurement at time instant $k=0,1,\ldots$, respectively. The functions $\bff_k:\real^{n_x}\mapsto\real^{n_x}$ and $\bfh_k:\real^{n_x}\mapsto\real^{n_z}$ are assumed known. The process noise $\bfw_k\in\real^{n_x}$ and measurement noise $\bfv_k\in\real^{n_z}$ are described by known \acp{pdf} $p(\bfw_k)$ and $p(\bfv_k)$. The initial state $\bfx_0$ is given by known \ac{pdf} $p(\bfx_0)$.
Both noises are assumed to be white, mutually independent, and independent of the initial state.

This paper assumes that the process noise is Gaussian\footnote{The notation $\calN\{\bfy;\bfm,\bfP\}$ denotes a Gaussian distribution of $\bfy$ with mean $\bfm$ and covariance matrix $\bfP$.} 
\begin{align}\label{eq:pdfw}
    p(\bfw_k)&=\calN\{\bfw_k;\bfnul_{n_x\times1},\bfQ\},
\end{align}
while some measurement channels are subject to outliers. Hence, the measurement noise is described by a product of zero-mean Student's-t (for the channels containing outliers) and Gaussian (for the remaining channels) distributions
    \begin{align}\label{eq:pdfv}
p(\bfv_k)&=p(\bfv_k^\tpdf)\cdot p(\bfv_k^\gpdf) ,
    \end{align}
where $\bfv_k=\begin{bsmallmatrix} \bfv_k^\gpdf\\\bfv_k^\tpdf\end{bsmallmatrix}$, $\bfv_k^\gpdf\in\real^{n_z^\gpdf}$ represents the Gaussian distributed elements of the measurement noise with 
\begin{align}
    p(\bfv_k^\gpdf)&=\calN\{\bfv_k^\gpdf;\bfnul_{n_z^\gpdf\times1},\bfR_k^\gpdf\}
\end{align}and $\bfv_k^\tpdf\in\real^{n_z^\tpdf}$ represents the Student's-t distributed elements
\begin{align}
    p(\bfv_k^\tpdf)&= \calT\{\bfv_k^\tpdf;\bfnul_{n_z^\tpdf\times1},\bS_k,\nu_k\}.
\end{align}
Here $\calT\{\bfy;\bfm,\bfSigma,\nu\}=\frac{\Gamma(\tfrac{\nu+2}{2})}{\Gamma(\tfrac{\nu}{2})}\frac{(1+\tfrac{(\bfy{-}\bfm)\T\bfSigma^{-1}(\bfy{-}\bfm)}{\nu})^{-\tfrac{\nu+2}{2}}}{(\nu\pi)^{n_y/2}|\bfSigma|^{(1/2)}}$, denotes the multivariate Student's-t density with $\Gamma$ denoting the Gamma function, $\bfm$ being the location parameter, and $\bfSigma$ being the scale parameter.
For $\nu{>}2$ the mean and variance of $\bfy$ are $\mean[\bfy]{=}\bfm$ and
$\cov[\bfy]=\frac{\nu}{\nu-2}\bfSigma$, respectively.

Given the  noise densities~\eqref{eq:pdfw} and \eqref{eq:pdfv}, the transition and measurement \acp{pdf} have the form
\begin{align}\label{eq:transition_gaussian}
    p(\bfx_{k+1}|\bfx_{k})=& \calN\{\bfx_{k+1};\bff_{k}(\bfx_k),\bfQ\} , \\
\label{eq:measurement_student}
    p(\bfz_{k+1}|\bfx_{k+1})=& 
    \calN\{\bfz_{k+1}^\gpdf;\bfh_{k+1}^\gpdf(\bfx_{k+1}),\bfR_{k+1}^\gpdf\}\nonumber\\
    &\calT\{\bfz_{k+1}^\tpdf;\bfh_{k+1}^\tpdf(\bfx_{k+1}),\bS_{k+1},\nu_{k+1}\} ,
\end{align}
such that $\bfz_{k+1}=\begin{bsmallmatrix} \bfz_{k+1}^\gpdf\\ \bfz_{k+1}^\tpdf \end{bsmallmatrix}$ and $\bfh_{k+1}(\bfx_{k+1})=\begin{bsmallmatrix}
    \bfh_{k+1}^\gpdf(\bfx_{k+1})\\\bfh_{k+1}^\tpdf(\bfx_{k+1})
\end{bsmallmatrix}$.

Assuming the knowledge of the model, the goal of Bayesian state estimation is to infer the posterior \ac{pdf} $p(\bfx_k|\bfz^k)$ of the state $\bfx_k$ given all the measurements available up to time~$k$, denoted as $\bfz^k\coloneqq[\bfz_1\T,\bfz_2\T,\cdots,\bfz_k\T]\T$.

\section{Structured Gaussian Mixture Filter}\label{sec:structured_gmf}
The general solution to the problem is provided by the \acp{brr} consisting of the Bayes equation~\eqref{eq:brrbayes} and the \ac{cke}~\eqref{eq:brrcke}
\begin{subequations}\label{eq:brr}
    \begin{align}\label{eq:brrbayes}        p(\bfx_{k+1}|\bfz^{k+1})&=\tfrac{p(\bfz_{k+1}|\bfx_{k+1})p(\bfx_{k+1}|\bfz^{k})}{p(\bfz_{k+1}|\bfz^{k})} , \\
        p(\bfx_{k+1}|\bfz^{k})&=\int p(\bfx_{k+1}|\bfx_{k})p(\bfx_{k}|\bfz^{k})\d\bfx_{k},\label{eq:brrcke}
    \end{align}
\end{subequations}
where $p(\bfz_{k+1}|\bfz^{k})$ is the measurement predictive \ac{pdf} given by $p(\bfz_{k+1}|\bfz^{k})=\int p(\bfz_{k+1}|\bfx_{k+1})p(\bfx_{k+1}|\bfz^{k})\d\bfx_{k+1}$.
The initial condition for the \acp{brr} is $p(\bfx_0|\bfz^0)=p(\bfx_0)$.
The calculation of the \acp{brr} thus involves alternating the filtering step~\eqref{eq:brrbayes} and the prediction step~\eqref{eq:brrcke}.

The fixed-structure \ac{gmf}~\cite{Straka25:Efficient} follows the \acp{brr} with the transition \ac{pdf} $p(\bfx_{k+1}|\bfx_k)$~\eqref{eq:transition_gaussian} decomposed\footnote{Note that the decomposition can be computed for other distributions of the transition \ac{pdf} such as Student's-t or generalized Gaussian~\cite{TiStraDu:23}.} as follows~\cite{TiStraDu:23} 
\begin{align}\label{eq:decomposition_predicted}
    p(\bfx_{k+1}&|\bfx_{k}) =\calN\{\bfx_{k+1};\bff_{k}(\bfx_k),\bfQ\}\nonumber\\
    &\approx \sum_{j=1}^{M_{k+1}}\omega^j\, \bfgamma^j[\bfx_{k+1};\bftheta^{\bfgamma,j}_{k+1}]\,\bfphi^j[\bff_{k}(\bfx_{k});\bftheta^{\bfphi,j}_k],
\end{align}
where $M_{k+1}$ is the rank of the (functional) decomposition and functions $\bfgamma^j$ and $\bfphi^j$ are given by a Gaussian \ac{pdf}
\begin{subequations}\label{eq:psg_decomposition}
\begin{align}\label{eq:psg_decomposition_xkpo}
    \bfgamma^j[\bfx_{k+1};\bftheta_{k+1}^{\bfgamma,j}]&=\calN\{\bfx_{k+1};\bfm^{\bfgamma,j}_{k+1},\bfP^{\bfgamma,j}\},\\
    \label{eq:psg_decomposition_fxk}
    \bfphi^j[\bff_{k}(\bfx_{k});\bftheta_{k}^{\bfphi,j}]&=\calN\{\bff_{k}(\bfx_{k});\bfm^{\bfphi,j}_k;\bfP^{\bfphi,j}\}.
\end{align}
\end{subequations}
The weights $\{\omega^j\}_{j=1}^{M_{k+1}}$ and the parameters $\{\bftheta^{\bfgamma,j}\}_{j=1}^{M_{k+1}}$ and $\{\bftheta^{\bfphi,j}\}_{j=1}^{M_{k+1}}$ are obtained by an optimization process off-line~\cite{TiStraDu:23}. The rank~$M_{k+1}$ depends on the size of the state space $\real^{n_x}$, which is to be covered by the decomposition, and on the denseness of the terms, which serves as a parameter for the optimization. 
The optimized parameters can be precomputed for several prescribed denseness levels, allowing for the decomposition~\eqref{eq:decomposition_predicted} to be readily constructed within a specified region of the state space for $\bfx_{k+1}$.
The decomposition is computed offline and therefore does not contribute to per-step filtering time. 
Its offline cost and storage, however, increase with the state-space region to be covered, the requested term denseness, and the state dimension, because these choices determine the decomposition rank $M$. 
Online complexity then grows approximately in proportion to the number of retained mixture terms. 
Consequently, high-dimensional problems may require coarser decompositions, structured construction, or additional pruning; efficient construction for high dimensions remains future work.

The fixed-structure \ac{gmf} involving the decomposition~\eqref{eq:decomposition_predicted} can be described using Algorithm~\ref{alg:gmf_psg}.
\begin{algorithm}
\caption{fixed-structure \ac{gmf}}\label{alg:gmf_psg}
\textbf{Step 1 (initialization):} Set $k=0$. Assume a posterior \ac{pdf}
\begin{align}\label{eq:initialization}
    p(\bfx_{k}|\bfz^{k})=\sum_{i=1}^{N_{k|k}}\alpha_{k|k}^i\,
    p^i(\bfx_k|\bfz^k).
\end{align}
\textbf{Step 2 (time update):} Calculate the \ac{cke}~\eqref{eq:brrcke} using the transition density decomposition~\eqref{eq:decomposition_predicted} as
\begin{multline}
    p(\bfx_{k+1}|\bfz^{k})=\int p(\bfx_{k+1}|\bfx_{k})p(\bfx_{k}|\bfz^{k})\d \bfx_k\\
    \approx \int\Bigg( \sum_{j=1}^{M_{k+1}}\omega^j_k\, \bfgamma^j[\bfx_{k+1};\bftheta^{\bfgamma,j}_{k+1}]\,\bfphi^j[\bff_{k}(\bfx_{k});\bftheta^{\bfphi,j}_k]\\
    \sum_{i=1}^{N_{k|k}}\alpha_{k|k}^i\,
    p^i(\bfx_k|\bfz^k)\Bigg)
    \d\bfx_k,
\end{multline}
which can be rewritten as 
\begin{align}\label{eq:structured}
    p(\bfx_{k+1}|\bfz^{k})\approx \sum_{j=1}^{M_{k+1}}\alpha_{k+1|k}^j\,\bfgamma^j[\bfx_{k+1};\bftheta^{\bfgamma,j}_{k+1}],
\end{align}
where
\begin{multline}\label{eq:gmf_psg_pred_weight}
    \alpha_{k+1|k}^j=\omega^j\!\!\sum_{i=1}^{N_{k|k}}\!\!\!\alpha_{k|k}^i
    \int\!\! \bfphi^j[\bff_{k}(\bfx_{k});\bftheta^{\bfphi,j}_k]
    p^i(\bfx_k|\bfz^k)
    \d\bfx_k.
\end{multline}
As the function $\bfgamma^j$ is of the form of a Gaussian \ac{pdf} (c.f.~\eqref{eq:psg_decomposition_xkpo}), the predictive PDF $p(\bfx_{k+1}|\bfz^{k})$ is a \ac{gm}
\begin{align}\label{eq:predictive_gm}
    p(\bfx_{k+1}|\bfz^{k})=\sum_{j=1}^{N_{k+1|k}}\alpha_{k+1|k}^j\,\calN\{\bfx_{k+1};\bfm^{\bfgamma,j}_{k+1},\bfP^{\bfgamma,j}\}.
\end{align}
From the predictive \ac{pdf}~\eqref{eq:predictive_gm}, the point estimate $\hbfx_{k+1|k}=\mean[\bfx_{k+1}|\bfz^k]$ and the covariance matrix $\bfP_{k+1|k}=\cov[\bfx_{k+1}|\bfz^k]$ can be computed if needed.

\textbf{Step 3 (measurement update):} Calculate the Bayes equation~\eqref{eq:brrbayes} 
using~\eqref{eq:predictive_gm} and \eqref{eq:measurement_student} leading to
\begin{align}\label{eq:posterior}
    p(\bfx_{k+1}|\bfz^{k+1})=\!\!\!\!\!\sum_{j=1}^{N_{k+1|k+1}}\!\!\!\!\!\alpha_{k+1|k+1}^j 
    p^j(\bfx_{k+1}|\bfz^{k+1}).
\end{align}
From the filtering \ac{pdf}~\eqref{eq:posterior}, the point estimate $\hbfx_{k+1|k+1}=\mean[\bfx_{k+1}|\bfz^{k+1}]$ and the covariance matrix $\bfP_{k+1|k+1}=\cov[\bfx_{k+1}|\bfz^{k+1}]$ can be computed.

\vspace{1em}
Continue with \textbf{Step~2}.
\end{algorithm}

\begin{figure}
    \centering
    \hspace{-2em}
    \includegraphics[width=0.9\linewidth]{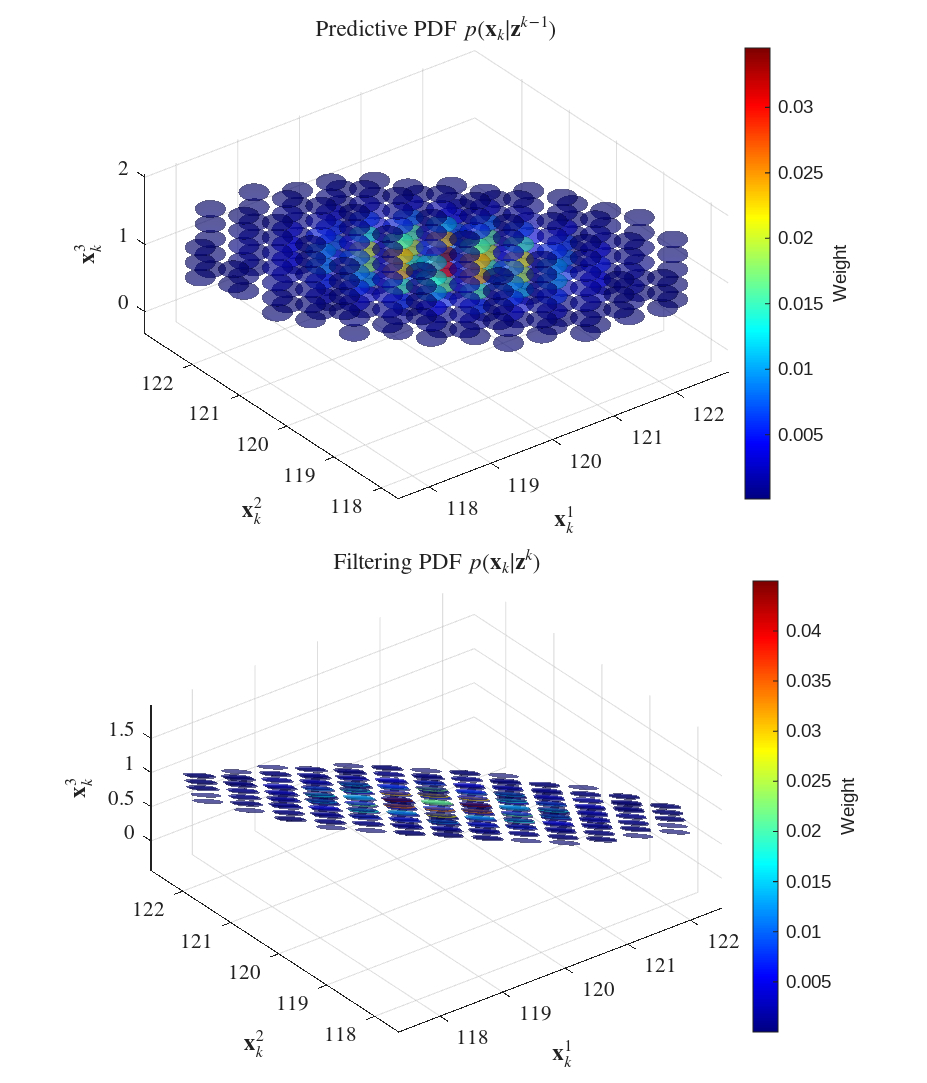}
    \caption{Illustration of the filtering and predictive \acp{pdf} of the structured \ac{gmf} with pruning for $n_x{=}3$, $\bfx_k{=}[\bfx_k^1,\bfx_k^2,\bfx_k^3]\T$, with the covariance represented through the ellipsoid sizes and weights represented by color.}
    \label{fig:illustration_gm_wp}
\end{figure}

\textbf{Remarks:} 
\begin{description}
    \item[(R1)]~The type of \acp{pdf} $p^i(\bfx_k|\bfz^k)$ in~\eqref{eq:initialization} and~\eqref{eq:posterior} depends on the measurement update, in particular on~$p(\bfz_k|\bfx_k)$ and the approximation used in~\eqref{eq:brrbayes}. It has no impact on the structure of the predictive \ac{pdf}~\eqref{eq:predictive_gm} since its structure is fixed through the decomposition, which is a \ac{gm}.
    \item[(R2)]~The information from the previous step stored in $p(\bfx_k|\bfz^k)$~\eqref{eq:initialization} is transformed into the weights $\alpha^j_{k+1|k}$ through \eqref{eq:gmf_psg_pred_weight}. 
The new measurement $\bfz_{k+1}$ then incurs an update of weights $\alpha_{k+1|k}^j$, and transformation of the \ac{pdf} $\calN\{\bfx_{k+1};\bfm^{\bfgamma,j}_{k+1},\bfP^{\bfgamma,j}\}$ to $p^j(\bfx_{k+1}|\bfz^{k+1})$. For $p^j(\bfx_{k+1}|\bfz^{k+1})$ being Gaussian, this transformation corresponds to a change of moments $\bfm^{\bfgamma,j}_{k+1}$ and $\bfP^{\bfgamma,j}$.
\item[(R3)]~The only parameter of the decomposition specified during the online run of the algorithm is the region of the state space that is to be covered by the decomposition. An efficient way is to calculate approximate moments of the filtering \ac{pdf} $p(\bfx_{k+1}|\bfz^{k+1})$ (e.g., by an \ac{ukf}) and use them to specify the region of the approximation.
\item [(R4)] If the rank of the decomposition $M_{k+1}$, given the calculated region, is too high, the terms with negligible weights $\alpha^j_{k+1|k}$ can be pruned since the terms associated with these weights will have negligible effect on the filtering \ac{pdf} $p(\bfx_{k+1}|\bfz^{k+1})$. This pruning is inspired by the fact that the term with a zero weight $\alpha^j_{k+1|k}$ in~\eqref{eq:structured}  will have no impact on~\eqref{eq:posterior} regardless of $\bfz_{k+1}$. 
\end{description}
The predictive and filtering \acp{pdf} of the fixed-structure \ac{gmf} after pruning of negligible terms are illustrated in
~\ref{fig:illustration_gm_wp} for the problem considered in the numerical illustration.

\section{Robust Measurement Update}\label{sec:measurement_update}
For convenience, assume all measurement noise elements are Student's-t distributed, i.e., $n_z = n_z^\tpdf$ and $n_z^\gpdf=0$. The case for combining Gaussian and Student's-t distributed measurement noise elements will be addressed later. For a Student's-t distributed noise, the measurement \ac{pdf} can be expressed using a hierarchical representation as
\begin{align}
\label{eq:measurement_pure_student}
p&(\bfz_{k+1}|\bfx_{k+1})=
\calT\{\bfz_{k+1};\bfh_{k+1}(\bfx_{k+1}),\bS_{k+1},\nu_{k+1}\}
\nonumber\\
&=
\int_0^\infty
\underbrace{
\calN\{\bfz_{k+1};\bfh_{k+1}(\bfx_{k+1}),\bS_{k+1}/\lambda_{k+1}\}
}_{p(\bfz_{k+1}|\bfx_{k+1},\lambda_{k+1})}
\,p(\lambda_{k+1})\,\mathrm{d}\lambda_{k+1},
\end{align}
with parameter $\lambda_{k+1}$ being Gamma distributed, $p(\lambda_{k+1})=\calG\{\lambda_{k+1};\tfrac{\nu_{k+1}}{2},\tfrac{\nu_{k+1}}{2}\}$.

To compute the term $p^j(\bfx_{k+1}|\bfz^{k+1})$ of the posterior \ac{gm}~\eqref{eq:posterior} from the term $\calN\{\bfx_{k+1};\bfm^{\bfgamma,j}_{k+1},\bfP^{\bfgamma,j}\}$ of the predictive \ac{gm}, we propose to use the \ac{vb} approach~\cite{NuArPiGu:15}. For convenience, the superscript $j$ and subscript $k{+}1$ will be dropped in the following derivation. The \ac{pdf} $p_\bfx(\bfx)$ denotes the $j$-th term of the predictive \ac{pdf} and $p_\bfx(\bfx|\bfz)$ denotes the $j$-th term of the posterior (filtering) \ac{pdf}.

The \ac{vb} approach seeks its approximation of the form
\begin{align}
    p(\bfx,\lambda|\bfz)\approx p_{\bfx}(\bfx)p_{\lambda}(\lambda)
\end{align}
by minimizing the Kullback-Leibler divergence of the true posterior from the factorized approximation~\cite{NuArPiGu:15}
\begin{align}\label{eq:kld}
    p^*_{\bfx}(\bfx)\,p^*_{\lambda}(\lambda)=\arg\min_{p_{\bfx},p_{\lambda}}D_\mathrm{KL}\Big(p_{\bfx}(\bfx)p_{\lambda}(\lambda)\|p(\bfx,\lambda|\bfz)\Big).
\end{align}
The analytic solution to~\eqref{eq:kld} can be obtained by a cyclic iteration of steps 
\begin{subequations}
    \begin{align}
        \log p_{\bfx}(\bfx) &= \mean_{p_{\lambda}}[\log p(\bfz,\bfx,\lambda)]+c_\bfx\\
        \log p_{\lambda}(\lambda) &= \mean_{p_{\bfx}}[\log p(\bfz,\bfx,\lambda)]+c_\lambda,
    \end{align}
\end{subequations}
where $p(\bfz,\bfx,\lambda)=p(\bfz|\bfx,\lambda)p(\bfx)p(\lambda)$, $c_\bfx$ and $c_\lambda$ are constants w.r.t. $\bfx$ and $\lambda$, respectively. This cyclic iteration is the core of the robust update of the proposed robust fixed-structure \ac{gmf}.

The algorithm of the measurement update of a single term $\calN\{\bfx_{k+1};\bfm_{k+1|k}^{\bfgamma,j},\bfP^{\bfgamma,j}\}$ of the predictive \ac{gm},  is described in Algorithm~\ref{alg:measurement_update}. The nonlinearity of $\bfh$ is addressed by \ac{ut}~\cite{JuUhl:04}, which is used to compute the predictive measurement mean $\hbfz_{k+1|k}=\mean[\bfz_{k+1}|\bfz^k]=\mean[\bfh_{k+1}(\bfx_{k+1})|\bfz^k]$, predictive covariance matrix of $\bfh_{k+1}(\bfx_{k+1})$ defined as $\bfP_{k+1|k}^{\bfh\bfh}=\var[\bfh_{k+1}(\bfx_{k+1})|\bfz^k]$, and cross covariance matrix $\bfP_{k+1|k}^{\bfx\bfh}=\cov[\bfx_{k+1},\bfh_{k+1}(\bfx_{k+1})|\bfz^k]$. 

\begin{algorithm}
\caption{Robust measurement update of a $j$-th \ac{gm} term}\label{alg:measurement_update}
\begin{algorithmic}
\Require $\calN\{\bfx_{k+1};\bfm_{k+1|k}^{\bfgamma,j},\bfP^{\bfgamma,j}\}$,~$\calG\{\lambda_{k+1};\tfrac{\nu_{k+1}}{2},\tfrac{\nu_{k+1}}{2}\}$,~$\bfz_{k+1}$
\Ensure $\calN\{\bfx_{k+1};\bfm_{k+1|k+1}^{\bfgamma,j},\bfP_{k+1|k+1}^{\bfgamma,j}\}$
\State Use \ac{ut} to compute $\hbfz_{k+1|k}^j$,  $\bfP_{k+1|k}^{\bfh\bfh,j}$, and  $\bfP_{k+1|k}^{\bfx\bfh,j}$.
\State Set $\bar{\lambda}_{k+1}=\mean_{p_{\lambda_{k+1}}}[\lambda_{k+1}]=1$.
    \Repeat
    \State Update $\calN\{\bfx_{k+1};\bfm_{k+1|k+1}^{\bfgamma,j},\bfP^{\bfgamma,j}_{k+1|k+1}\}$ given $p_\lambda(\lambda)$ as
    \State 
    \begin{align*}   \bfP_{k+1|k}^{\bfz\bfz,j}&=\bfP_{k+1|k}^{\bfh\bfh,j}+\bfSigma_{k+1}/\bar{\lambda}_{k+1}\\
    \bfK_{k+1}^j &=  \bfP_{k+1|k}^{\bfx\bfh,j}(\bfP_{k+1|k}^{\bfz\bfz,j})^{-1}\\
    \bfm_{k+1|k+1}^{\bfgamma,j}&=\bfm_{k+1|k}^{\bfgamma,j}+\bfK_{k+1}^j(\bfz_{k+1}-\hbfz_{k+1|k}^j)\\
    \bfP_{k+1|k+1}^{\bfgamma,j}&=\bfP^{\bfgamma,j}-\bfK_{k+1}^j(\bfP_{k+1|k}^{\bfx\bfh,j})\T
    \end{align*}
    \State Update $p_{\lambda}(\lambda)=\calG\{\lambda;\tfrac{\nu_{k+1}+n_z}{2},\tfrac{\nu_{k+1}+\bfXi_{k+1}}{2}\}$
    \State $\qquad \bfXi_{k+1}=(\bfz_{k+1}{-}\hbfz_{k+1|k}^j)\T(\bfP_{k+1|k}^{\bfz\bfz,j})^{-1}(\bfz_{k+1}{-}\hbfz_{k+1|k}^j)$
\State $\qquad \bar{\lambda}_{k+1}=\mean_{p_{\lambda_{k+1}}}[\lambda_{k+1}]=\tfrac{\nu_{k+1}+n_z}{\nu_{k+1}+\bfXi_{k+1}}$
    \Until{converged}
    \end{algorithmic}
\end{algorithm}

\textbf{Algorithm for the measurement noise given by a combination of Gaussian \ac{pdf} and Student's-t \ac{pdf}~\eqref{eq:pdfv}:}
The fixed-structure \ac{gmf} with robust measurement update proposed in this paper is given by the Algorithm~\ref{alg:gmf_psg} with the measurement update step specified as follows: When the measurement noise consists of both Gaussian and Student's-t distributed elements~\eqref{eq:pdfv}, two measurement update steps must be performed sequentially for each term of the \ac{gm}. 
For the exact Bayesian update, the Gaussian and Student's-t likelihood factors commute. In the proposed implementation, however, each sub-update is approximate (\ac{ukf} moment approximation and \ac{vb} factorization), so reversing their order can, in general, lead to a different Gaussian approximation. We use the Gaussian channel first and the robust Student's-t channel second consistently throughout the experiments. A systematic comparison of the two orders is outside the present scope.

\begin{align}
    p(\bfz_{k+1}|\bfx_{k+1})=& 
    \calN\{\bfz_{k+1}^\gpdf;\bfh_{k+1}^\gpdf(\bfx_{k+1}),\bfR_{k+1}^\gpdf\}
\end{align}
followed by the measurement update step for the Student's-t elements given by Algorithm~\ref{alg:measurement_update}, considering the \ac{pdf}
\begin{align}
    p(\bfz_{k+1}|\bfx_{k+1})=\calT\{\bfz_{k+1}^\tpdf;\bfh_{k+1}^\tpdf(\bfx_{k+1}),\bS_{k+1},\nu_{k+1}\}.
\end{align}
Application of the above \ac{vb}-based steps leads to the filtering \ac{pdf} in the \ac{gm} form
\begin{multline}\label{eq:posterior_gaussian}
    p(\bfx_{k+1}|\bfz^{k+1})=\sum_{j=1}^{N_{k+1|k+1}} \alpha_{k+1|k+1}^j \\
     \calN\{\bfx_{k+1};\bfm^{\bfx,j}_{k+1|k+1},\bfSigma^{\bfx,j}_{k+1|k+1}\}.
\end{multline}
It only remains to specify the measurement update of the weights $\alpha_{k+1|k}^j$ using the observed measurement $\bfz_{k+1}$
\begin{multline}
\alpha_{k+1|k+1}^j \propto \alpha_{k+1|k}^j\, \calN\{\bfz_{k+1}^\gpdf;\hbfz_{k+1|k}^{\gpdf},\bfP_{k+1|k}^{\bfh\bfh,j,\gpdf}+\bfR_{k+1}^{\gpdf}\}\\ \calT\{\bfz_{k+1}^\tpdf;\hbfz_{k+1|k}^{\tpdf},\bfSigma_{k+1},\bar{\lambda}_{k+1}^{j}\}
,
\end{multline}
where the measurement prediction mean $\hbfz_{k+1|k}^{j,\gpdf}$ and covariance matrix $\bfP_{k+1|k}^{\bfh\bfh,j,\gpdf}$ are obtained by the \ac{ut} in the measurement update for the Gaussian elements and the measurement prediction mean $\hbfz_{k+1|k}^{\tpdf}$ and the mean of the Gamma distribution $\bar{\lambda}_{k+1}^{j}$ are produced by Algorithm~\ref{alg:measurement_update} for the $j$-th term.


\section{Numerical Illustration}\label{sec:numerical illustration}
The performance of the proposed algorithm is illustrated using a numerical example of object tracking. Consider an object moving in a two-dimensional plane with its position at time $t_k$ denoted by $[\xpos_k\, \ypos_k]\  [m]$.
The object's position is observed by a radar sensor, which provides the range $r_k\,[m]$ and bearing $\rho_k\ [rad]$. 
The range is subject to a zero-mean Gaussian noise, and the bearing is subject to a noise with outliers \emph{simulated} by a mixture of Gaussian \acs{pdf} $p(\rho_k)=0.9\,\calN\{\rho_k;0,2\times 10^{-4}\}+0.1\,\calN\{\rho_k;0,77\times 10^{-4}\}$.
In addition, the bearing measurement is affected by an unknown bias denoted by $b_k$.
Thus, the target motion is two-dimensional, whereas the augmented estimation state is three-dimensional due to the additional bearing-bias component.
Therefore, the state $\bfx_k$ of the object model is defined as $\bfx_k=[\xpos_k,\ypos_k,b_k]\T$.
Its dynamic equation is of the form:
\begin{align}
\underbrace{
    \begin{bsmallmatrix}
        \xpos_{k+1}\\\ypos_{k+1}\\b_{k+1}
    \end{bsmallmatrix}}_{\bfx_{k+1}}    
    =
    \underbrace{
    \begin{bsmallmatrix}
        1.1&0&0\\
        0&1.1&0\\
        0&0& \exp(-1/\beta)T_s
    \end{bsmallmatrix}}_{\bfF}    
\underbrace{
    \begin{bsmallmatrix}
        \xpos_{k}\\\ypos_{k}\\b_{k}
    \end{bsmallmatrix}}_{\bfx_{k}}
    +\bfw_k,
\end{align}
where the process noise $\bfw$ is white zero-mean Gaussian with a known covariance matrix $\bfQ=10^{-1}\begin{bsmallmatrix}5&0&0\\0&5&0\\0&0&0.5\end{bsmallmatrix}$. The parameter $\beta$ of the bias process is set to $\beta=100\, s$, and the sampling period is $T_s=1\, s$. 
The measurement equation is given by
\begin{align}
\underbrace{\begin{bsmallmatrix}
        r_k \\ \rho_k
    \end{bsmallmatrix}}_{\bfz_k}
    =
    \underbrace{
    \begin{bsmallmatrix}
        \sqrt{\xpos_k^2+\ypos_k^2}\\
        \atantwo(\ypos_k,\xpos_k)+b_k
    \end{bsmallmatrix}}_{\bfh(\bfx_k)}
    +
    \underbrace{\begin{bmatrix}
        v^r_k\\v^\rho_k
    \end{bmatrix}}_{\bfv_k}.
\end{align}
The range noise is zero-mean Gaussian, $p(v^r_k)=\calN\{v^r_k;0,R\}$ with $R=0.1^2$. The bearing noise is \emph{modelled} as zero-mean Student's-t distributed $p(v^\rho_k)=\calT\{v^\rho_k,0,\Sigma,3\}$. Two cases of $\Sigma$ are considered, $\Sigma_a=(1\tfrac{\pi}{180})^2$ and $\Sigma_b=(2\tfrac{\pi}{180})^2$, where the second case is more challenging compared to the first case since the measurement \ac{pdf} $p(\bfz_k|\bfx_k)$ exhibits a "banana shape". This shape arises from a small uncertainty in range and a larger one in bearing. Note that the \ac{gm} used for the \emph{simulation} of the bearing noise with outliers is chosen such that its variance is equal to the variance of the Student's-t distribution used as a \emph{model for the filters} (The values correspond to $\Sigma_a$. For $\Sigma_b$, the GM variances are appropriately increased.). The process and measurement noises are mutually independent and independent of the initial condition $\bfx_0$, which is Gaussian with mean $[100,\,100,\,1]\T$ and covariance matrix $\cov[\bfx_0]=0.1\cdot\bfI_3$.

For the comparison, three filters are considered:
\begin{itemize}
    \item the proposed fixed-structure \ac{gmf} with robust measurement update with pruning of insignificant terms (rfsGMF) with threshold parameter $0.1$, maximum nuber of iterations $20$
    \item the \ac{ukf} with the parameter $\kappa=0$
    \item the \ac{bpf} with $10^5$ samples and multinomial resampling.    
\end{itemize}
The \ac{ukf} represents a \ac{gaf} providing only the conditional mean $\mean[\bfx_k|\bfz^k]$ and covariance matrix $\cov[\bfx_k|\bfz^k]$. It operates only with the first two moments of the measurement noise. The \ac{bpf} can work with arbitrary measurement noise \ac{pdf}. Note that the Student's-t filters were not considered, as they assume a Student's-t distribution of both process and measurement noises, which is not the case in the considered example.

The performance of the filters was measured using the \ac{rmse} defined as
\begin{align}
    \rmse_k = \left(\tfrac{1}{M}\sum_{\ell=1}^{M}(\tx_{k|k}(\ell))\strut^2\right)^{1/2}
\end{align}
based on $M=10^3$ \ac{mc} simulations with $\tx_{k|k}(\ell)\coloneq x_k(\ell)-\hx_{k|k}(\ell)$ being the estimate error,  $x_k(\ell)$ being the true state at $\ell$-th \ac{mc} simulation and $\hx_{k|k}(\ell)$ being its filtering estimate. 
To assess higher-order information provided by the filters, the \ac{anees}~\cite{LiZha:06a} defined by
\begin{align}
    \anees_k = \frac{1}{n_x}\frac{1}{M}\sum_{\ell=1}^{M}\tbfx_{k|k}(\ell)\T\,\cov[\bfx_k|\bfz^k]^{-1}\,    \tbfx_{k|k}(\ell)
\end{align}
is used. \ac{anees} assesses the consistency of the estimator, i.e., alignment of the conditional covariance matrix $\cov[\bfx_k|\bfz^k]$ and the estimate error $\tbfx_{k|k}$. This value should be close to one. Higher or lower \ac{anees} values mean that the estimator is too optimistic or too pessimistic, respectively.

The values of \ac{rmse} and \ac{anees} for filtering estimates $\hbfx_{k|k}=\mean[\bfx_k|\bfz^k]$, $\cov[\bfx_k|\bfz^k]$ 
averaged over $k=1,\ldots,30$ are given in Table~\ref{tab:results}.
\begin{table}[ht]
\renewcommand{\arraystretch}{1.2}
    \centering
    \caption{Time average of \ac{rmse} and \ac{anees}.}
    \label{tab:results}
    \begin{tabular}{lccccccc}
    \toprule
    &&\multicolumn{3}{c}{$\Sigma_a=(1\tfrac{\pi}{180})^2$}&\multicolumn{3}{c}{$\Sigma_b=(2\tfrac{\pi}{180})^2$}\\
    & & \ac{ukf} & \ac{bpf} & rfsGMF & \ac{ukf} & \ac{bpf} & rfsGMF \\\midrule
    \multirow{3}{3em}{\ac{rmse}} & $\xpos$ & 8.435 & 6.906 & 6.321 & 9.510 & 10.459 & 8.514 \\
         & $\ypos$ & 8.309 & 6.752 & 6.271 & 9.606 & 10.359 & 8.615 \\
         & $b$ & 0.794 & 0.782 & 0.794 & 0.664 & 0.754 & 0.661 \\
     \ac{anees} & & 2.584 & 2.756 & 1.901 & 2.308 & 3.599 & 2.132 \\\bottomrule
    \end{tabular}
\end{table}
The table indicates that, in terms of \ac{rmse}, the rfsGMF provides the best performance for both cases $\Sigma_a$ and $\Sigma_b$. For the less challenging $\Sigma_a$, the worst performance is achieved by \ac{ukf} assuming a Gaussian posterior \ac{pdf}, and \ac{bpf} is slightly worse than rfsGMF. For the more challenging $\Sigma_b$, the performance of \ac{bpf} worsens significantly and is surpassed by the \ac{ukf} despite the large number of samples of \ac{bpf}. The issue is caused by the "banana shape" of the measurement \ac{pdf}, leading to many samples having negligible weights and thus not contributing to the state estimate.

The \ac{anees} values are above one for all three estimators, indicating that all methods are somewhat optimistic in their covariance assessment. 
The rfsGMF nevertheless provides the values closest to one in both scenarios. 
A more detailed investigation of the source of this residual inconsistency is beyond the scope of the present numerical study.

\section{Conclusion}\label{sec:conclusion}
The paper addressed Bayesian state estimation for nonlinear stochastic systems with measurement channels influenced by outliers. 
A robust measurement-update variant of the fixed-structure \ac{gmf} was proposed. 
The fixed structure avoids random sampling effects and uncontrolled mixture growth; its computational benefit depends on the selected decomposition rank and pruning level.
Robustness to outliers is achieved by modeling measurement noise with a Student's-t distribution and employing the variational Bayes method during the measurement update of individual Gaussian mixture terms. 
Numerical results demonstrated that the proposed filter consistently outperforms both the Gaussian-assumed unscented Kalman filter and the bootstrap particle filter. 
Future work will focus on efficiently constructing the fixed Gaussian mixture structure for high-dimensional problems.
\bibliographystyle{IEEEtran}

\end{document}